\documentclass[twocolumn,aps,pra,amsmath,superscriptaddress,notitlepage,longbibliography]{revtex4-1}
\usepackage{amsmath,amssymb,amsfonts,bm}
\usepackage{graphicx}
\usepackage{dcolumn}
\usepackage{mathrsfs}
\usepackage{bbold}
\usepackage{dsfont}
\usepackage{dcolumn}
\usepackage[colorlinks=true,linkcolor=blue,citecolor=blue, urlcolor=blue,bookmarks=false]{hyperref}
\usepackage{changes}

\begin{document}


\title{Tunable corner states in topological insulators with long-range hoppings and diverse shapes}

\author{Fang Qin}
\email{qinfang@just.edu.cn}
\affiliation{School of Science, Jiangsu University of Science and Technology, Zhenjiang, Jiangsu 212100, China}

\author{Rui Chen}
\email{chenr@hubu.edu.cn}
\affiliation{Department of Physics, Hubei University, Wuhan, Hubei 430062, China}

\begin{abstract}
In this work we develop a theoretical framework for the control of corner modes in higher-order topological insulators (HOTIs) featuring long-range hoppings and diverse geometries, enabling precise tunability of their spatial positions. First, we demonstrate that the locations of corner states can be finely tuned by varying long-range hoppings in a circular HOTI, as revealed by a detailed edge theory analysis and the condition of vanishing Dirac mass. Moreover, we show that long-range hoppings in different directions (e.g., $x$ and $y$) have distinct effects on the positioning of corner states. Second, we investigate HOTIs with various polygonal geometries and find that the presence and location of corner modes depend sensitively on the shape. In particular, a corner hosts a localized mode if the Dirac masses of its two adjacent edges have opposite signs, while no corner mode emerges if the masses share the same sign. Our findings offer a versatile approach for the controlled manipulation of corner modes in HOTIs, opening avenues for the design and implementation of higher-order topological materials.
\end{abstract}
\maketitle

\section{Introduction}

In two-dimensional (2D) topological materials, the bulk-edge correspondence is a key feature used to characterize topological insulators, where a gapped insulating bulk is associated with robust conducting states at the edges~\cite{kane2005z,kane2005quantum,hasan2010colloquium,moore2010birth,bernevig2006quantum,qi2006topological,qi2011topological}. These robust edge states are protected by time-reversal symmetry (TRS). For instance, the quantum spin Hall insulator, first proposed by Kane and Mele~\cite{kane2005z,kane2005quantum}, is a $\mathbb{Z}_{2}$ topological insulator protected by TRS. The Bernevig-Hughes-Zhang (BHZ) model~\cite{bernevig2006quantum} provides a framework for describing the quantum spin Hall insulator, where helical edge states are likewise protected by TRS. In contrast, the Qi-Wu-Zhang model~\cite{qi2006topological,qi2011topological}, also known as the half-BHZ model, serves as a fundamental model for studying topological insulators in the absence of TRS.

In 2D Dirac materials, higher-order topological insulators (HOTIs) host additional topologically protected localized states at the corners, beyond the states found at the edges~\cite{benalcazar2017quantized,benalcazar2017electric,song2017d,langbehn2017reflection,khalaf2018higher,yan2018majorana,agarwala2020higher,schindler2020dirac,xue2021higher,ghosh2024generation,yang2024higher,li2020topological,roy2019antiunitary,tao2023average}.
It is well known that a $D$-dimensional topological insulator features gapless boundary states on its $(D\!-\!1)$-dimensional edges. In contrast, a $D$-dimensional HOTI is characterized by gapped $(D\!-\!1)$-dimensional boundaries but hosts gapless states at even lower-dimensional boundaries, such as corners.
The realization of 2D HOTIs relies on specific spatial symmetries, such as mirror, rotation, and inversion symmetries, which play a crucial role in defining the topological invariants~\cite{li2020topological,roy2019antiunitary,tao2023average} that characterize the corner states in 2D HOTIs ~\cite{song2017d,langbehn2017reflection,khalaf2018higher}. These symmetries ensure the robustness of the corner states in 2D HOTIs, protecting the localized states from external perturbations.
Experimental realizations of second-order topological corner states have been observed in a variety of systems, including electrical circuits~\cite{imhof2018topolectrical,peterson2018a,bao2019topoelectrical,ezawa2019braiding,serra2019observation,kempkes2019robust,zhang2021experimental,lv2021realization,dong2021topolectric}, acoustic crystals~\cite{ni2019observation,qi2020acoustic,xue2019acoustic,xue2019realization,xue2020observation,gao2021non,li2021measurement}, and photonic crystals~\cite{serra2018observation,li2020higher,xie2019visualization,chen2019direct,mittal2019photonic,ei2019corner,kim2020recent,schulz2022photonic,mandal2024photonic}.
Higher-order topological insulators have also been realized in quasicrystals~\cite{varjas2019topological,chen2020higher,peng2021higher,traverso2022role,chen2023quasicrystalline,shi2025non} and in electronic systems such as bismuth fractal nanostructures~\cite{canyellas2024topolectrical}. Numerous candidate materials and structures have been theoretically predicted to exhibit higher-order topological phases. These include 2D hexagonal-lattice materials~\cite{qian2021second,pan2022two}, Kekul\'{e}-lattice graphdiyne~\cite{mu2022kekule}, transition-metal dichalcogenides~\cite{qian2022c,zeng2021multiorbital}, breathing kagome and pyrochlore lattices~\cite{wakao2020higher,ezawa2018higher}, black phosphorene~\cite{ezawa2018minimal}, graphdiyne~\cite{sheng2019two}, twisted bilayer graphene~\cite{park2019higher,spurrier2020kane}, and twisted moir\'{e} superlattice~\cite{liu2021higher}. For three-dimensional HOTIs, proposed candidates include bismuth~\cite{schindler2018higher1,zhang2023scanning}, SnTe~\cite{schindler2018higher2}, EuIn$_2$As$_2$~\cite{xu2019higher}, EuSn$_2$As$_2$~\cite{li2019dirac}, MnBi$_{2n}$Te$_{3n+1}$~\cite{li2019dirac,zhang2020mobius}, Bi$_{2-x}$Sm$_{x}$Se$_{3}$~\cite{yue2019symmetry}, and CrI$_{3}$/Bi$_{2}$Se$_{3}$/MnBi$_{2}$Se$_{4}$ heterostructures~\cite{hou2020axion}.
Recent theoretical studies have shown that higher-order topological phase transitions in Chern insulators can be realized by coupling two Chern insulators with opposite Chern numbers~\cite{mandal2024photonic,liu2024engineering,liu2025engineering}. Moreover, the positions of the corner states in twisted bilayer Chern insulators can be tuned through interlayer coupling~\cite{miao2025tunable}. The tunability of cornerlike modes in generalized quadrupole topological insulators has also been explored~\cite{chen2024tunable}, as well as the manipulation of higher-order topological states using altermagnets~\cite{li2024creation}. Additionally, the manipulation of corner states has been achieved using magnetic fields~\cite{zhu2018tunable}, electric fields~\cite{zhang2020all}, and sublattice degrees of freedom~\cite{kheirkhah2022corner,zhu2023sublattice}.

In contrast to previous works~\cite{yan2018majorana,agarwala2020higher}, where corner states are induced solely by nearest-neighbor hoppings, our model incorporates both nearest-neighbor and long-range hoppings, enabling greater flexibility and control over the emergence and properties of corner modes. In this work we establish a comprehensive theoretical framework for the manipulation of corner modes in HOTIs with long-range hoppings and diverse geometric configurations. We demonstrate that in circular HOTIs, the spatial locations of corner states can be finely tuned by adjusting the range of long-range hoppings. This tunability is quantitatively supported by a rigorous edge theory analysis, where the emergence of corner modes is governed by the condition of vanishing Dirac mass. Additionally, we show that long-range hoppings in different directions, such as along the $x$ and $y$ axes, have distinct and anisotropic effects on the positioning of corner modes. Extending our analysis to HOTIs with various polygonal shapes, we find that the geometry plays a crucial role in determining the presence and location of corner states. This geometric dependence can likewise be understood through the edge theory: A corner supports a localized mode when the Dirac masses of its two adjacent edges have opposite signs, whereas no corner mode appears if the masses share the same sign. These findings highlight a powerful mechanism for engineering and controlling corner states, offering possibilities for designing reconfigurable higher-order topological materials.

The paper is organized as follows. In Sec.~\ref{2} we introduce the tight-binding model Hamiltonian. In Sec.~\ref{3} we present numerical results and provide a detailed discussion of tunable corner states induced by long-range hoppings. Section~\ref{4} is devoted to deriving the effective edge model and determining the Dirac mass within the edge theory. Section~\ref{5} focuses on geometry-dependent tunable corner states based on numerical analysis. In Sec.~\ref{6} we discuss the potential experimental realization of our model. We summarize our findings in Sec.~\ref{7}.

\section{Model}\label{2}

We consider the real-space tight-binding model on a square lattice as
\begin{eqnarray}
\hat{H}_{\rm TB}^{}&\!\!=\!\!&\sum_{j_{x},j_{y}}\left[\hat{C}_{j_{x},j_{y}}^{\dagger}\hat{h}_{0}\hat{C}_{j_{x},j_{y}}^{} \right.\nonumber\\
&&\left.\!+\!\left(\!\hat{C}_{j_{x},j_{y}}^{\dagger}\hat{t}_{x}\hat{C}_{j_{x}+1,j_{y}}^{} \!+\! \hat{C}_{j_{x},j_{y}}^{\dagger}\hat{t}_{y}\hat{C}_{j_{x},j_{y}+1}^{}\!\right) \!+\! {\rm H.c.} \right.\nonumber\\
&&\left.\!+\!\left(\!\hat{C}_{j_{x},j_{y}}^{\dagger}\hat{t}_{x}^{(n)}\hat{C}_{j_{x}+n,j_{y}}^{} \!+\! \hat{C}_{j_{x},j_{y}}^{\dagger}\hat{t}_{y}^{(m)}\hat{C}_{j_{x},j_{y}+m}^{}\!\right) \!+\! {\rm H.c.} \!\right]\!,\label{eq:Hr_TB} \nonumber\\
\end{eqnarray} where $\hat{C}_{j_{x},j_{y}}^{}\!=\!(\hat{c}_{j_{x},j_{y},+,\uparrow}^{}, \hat{c}_{j_{x},j_{y},-,\uparrow}^{}, \hat{c}_{j_{x},j_{y},+,\downarrow}^{}, \hat{c}_{j_{x},j_{y},-,\downarrow}^{})^{T}$; $\hat{c}_{j_{x},j_{y},\pm,\uparrow/\downarrow}^{\dag}$ ($\hat{c}_{j_{x},j_{y},\pm,\uparrow/\downarrow}^{}$) is the electron creation (annihilation) operator for an electron with orbit $\pm$ and pseudospin up/down $(\uparrow\!/\!\downarrow)$ at the site $(j_{x},j_{y})$; ${\rm H.c.}$ denotes Hermitian conjugate; $n,m\in\mathbb{Z}$; and
\begin{eqnarray}
&&\hat{h}_{0}\!=\!\left( M\!-\!\frac{4B}{a^2} \right)\!\sigma_{0}\otimes\tau_{z}, \label{eq:h0}\\
&&\hat{t}_{x}\!=\!\frac{B}{a^2}\sigma_{0}\otimes\tau_{z} \!-\! i\frac{A}{2a}\sigma_{z}\otimes\tau_{x},\label{eq:tx}\\
&&\hat{t}_{y}\!=\!\frac{B}{a^2}\sigma_{0}\otimes\tau_{z} \!-\! i\frac{A}{2a}\sigma_{0}\otimes\tau_{y}, \label{eq:ty}\\
&&\hat{t}_{x}^{(n)}\!=\!-\frac{g}{a^{2}}\sigma_{x}\otimes\tau_{x},\label{eq:txn}\\
&&\hat{t}_{y}^{(m)}\!=\!\frac{g}{a^{2}}\sigma_{x}\otimes\tau_{x}.\label{eq:tym}
\end{eqnarray}
Here $\hat{t}_{x}^{(n)}$ and $\hat{t}_{y}^{(m)}$ represent the long-range hopping terms in the $x$ and $y$ directions, respectively, with $n\!>\!1$ and $m\!>\!1$. Specifically, $\hat{t}_{x}^{(n)}$ describes the hopping between site $(j_{x},j_{y})$ and site $(j_{x}\!+\!n,j_{y})$, while $\hat{t}_{y}^{(m)}$ describes the hopping between site $(j_{x},j_{y})$ and site $(j_{x}, j_{y}\!+\!m)$. For example, when $n\!=\!m\!=\!2$, these terms describe next-nearest-neighbor hoppings in the $x$ and $y$ directions, respectively. The matrices $\sigma_{x,y,z}$ and $\tau_{x,y,z}$ are Pauli matrices representing the spin and orbital degrees of freedom, while $\sigma_{0}$ is the $2\times2$ identity matrix. The parameter $a$ denotes the lattice constant and $A$, $B$, $M$, and $g$ are model parameters.

To capture the main physics of the long-range hopping terms $\hat{t}_{x}^{(n)}$ and $\hat{t}_{y}^{(m)}$ in Eq.~\eqref{eq:Hr_TB}, we further derive the corresponding momentum-space tight-binding model in the basis $(\hat{c}_{{\bf k},+,\uparrow}^{}, \hat{c}_{{\bf k},-,\uparrow}^{}, \hat{c}_{{\bf k},+,\downarrow}^{}, \hat{c}_{{\bf k},-,\downarrow}^{})^{T}$ as
\begin{eqnarray}
\hat{\cal H}_{\rm TB}^{}({\bf k})&\!\!=\!\!&M({\bf k})\sigma_{0}\otimes\tau_{z} \!+\! A_{x}\sin(k_{x}a)\sigma_{z}\otimes\tau_{x} \nonumber\\
&&\!+ A_{y}\sin(k_{y}a)\sigma_{0}\otimes\tau_{y} \!+\! \hat{\cal H}\rq{}_{\rm TB}({\bf k}),\label{eq:Hk_TB}
\end{eqnarray} where $\hat{c}_{{\bf k},\pm,\uparrow/\downarrow}^{\dag}$ ($\hat{c}_{{\bf k},\pm,\uparrow/\downarrow}^{}$) is the electron creation (annihilation) operator in momentum space, ${\bf k}\!=\!(k_{x},k_{y})$, $M({\bf k})\!=\!M_{0}\!+\!t_{x}\cos(k_{x}a)\!+\!t_{y}\cos(k_{y}a)$, $M_{0}\!=\!M\!-\!4B/a^2$, $t_{x}\!=\!t_{y}\!=\!2B/a^2$, $A_{x}\!=\!A_{y}\!=\!A/a$, $\hat{\cal H}\rq{}_{\rm TB}({\bf k})\!=\!\Delta_{0}[\cos(nk_{x}a)\!-\!\cos(mk_{y}a)]\sigma_{x}\otimes\tau_{x}$ represents the contribution from the long-range hopping terms $\hat{t}_{x}^{(n)}$ and $\hat{t}_{y}^{(m)}$ in Eq.~\eqref{eq:Hr_TB}, and $\Delta_{0}\!=\!-(2g/a^{2})$.
In particular, we will find that $\hat{\cal H}\rq{}_{\rm TB}({\bf k})$ can gap out the helical edge states, leading to the emergence of corner modes at the interface between domains of opposite Dirac mass.
Additionally, Ref.~\cite{yan2018majorana} chose $\hat{\cal H}\rq{}_{\rm TB}({\bf k})\!=\!\Delta_{0}[\cos(k_{x}a)\!-\!\cos(k_{y}a)]\sigma_{y}\otimes\tau_{y}$ and Ref.~\cite{agarwala2020higher} chose $\hat{\cal H}\rq{}_{\rm TB}({\bf k})\!=\!\Delta_{0}[\cos(k_{x}a)\!-\!\cos(k_{y}a)]\sigma_{x}\otimes\tau_{x}$. Physically, both works~\cite{yan2018majorana,agarwala2020higher} created corner states only induced by nearest-neighbor hopping. However, our formalism can create corner states induced by both nearest-neighbor and long-range hoppings.

For our following numerical calculations, a detailed derivation of the matrix representation of the real-space tight-binding model Hamiltonian with the open boundary conditions in both the $x$ and $y$ directions is provided in Sec. SI of the Supplemental Material~\cite{supp}.

\section{Tunable corner states with long-range hoppings}\label{3}

We present numerical results and corresponding discussions about the energy levels, tunable corner states, and Dirac mass in circular topological insulators with long-range hoppings varied in the $x$ and $y$ directions. For all numerical calculations, we set the parameters as $A\!=\!1$, $B\!=\!1$, $M\!=\!1$, $g\!=\!-0.25$, and lattice constant $a\!=\!1$.

\begin{figure*}[htpb]
\centering
\includegraphics[width=0.65\textwidth]{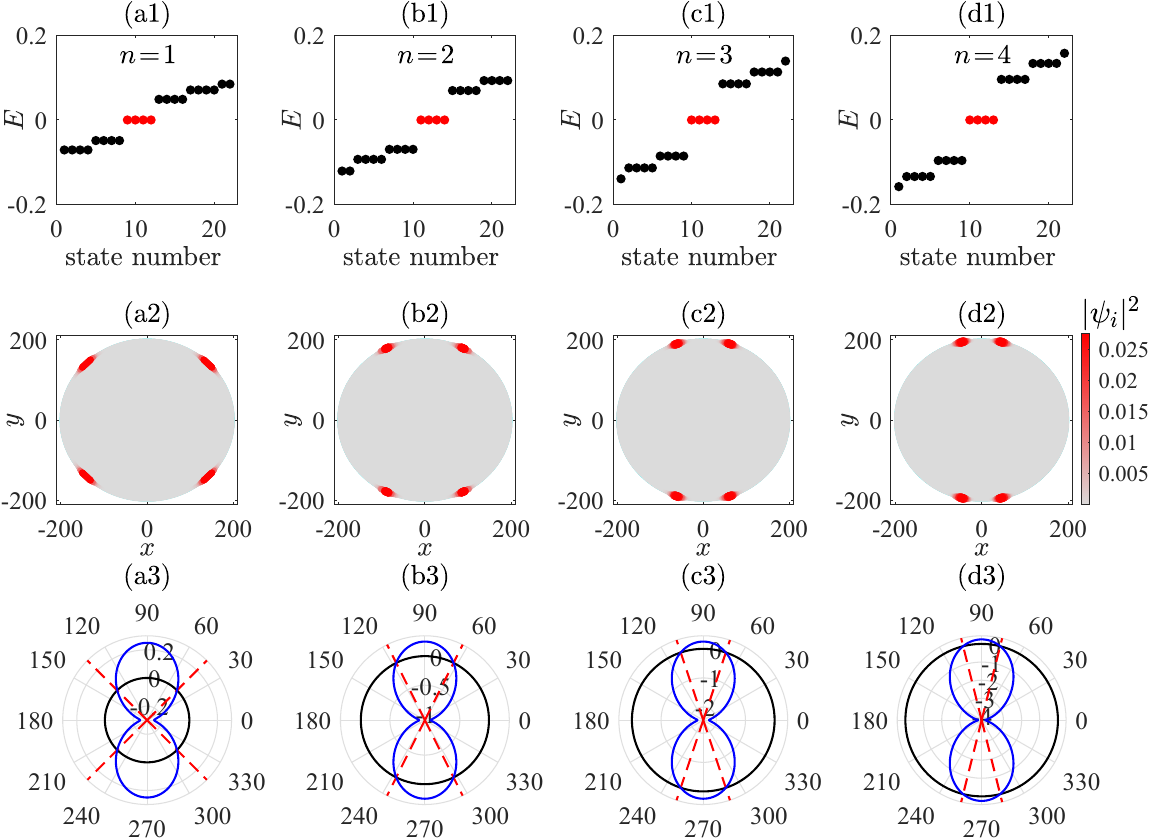}
\caption{Energy levels, corner states, and Dirac mass in a circular topological insulator with varying long-range hoppings in the $x$ direction, i.e., by changing $n\!=\!1,2,3,4$ while keeping $m\!=\!1$ fixed. (a1-d1) Energy levels. Here the red dots correspond to zero-energy modes. (a2-d2) Probability distribution of the corner states, highlighted in red. (a3-d3) Dirac mass (blue curve) [Eq.~\eqref{eq:Dirac_mass}] as a function of the normal angle $\alpha$ in the polar coordinate chart. Here, the black circle describes the zeros of Dirac mass along different polar angles, while the two red dashed lines indicate the normal directions associated with the corner modes. The parameters are $m\!=\!1$, $A\!=\!1$, $B\!=\!1$, $M\!=\!1$, $g\!=\!-0.25$, $L_{x}\!=\!L_{y}\!=\!401a$ with $a\!=\!1$.}
\label{fig:E_OBCxy_n_m1_circle_together}
\end{figure*}

\subsection{Case of $n\!=\!1,2,3,4$ with $m\!=\!1$}\label{3_1}

We compute and plot the energy levels, probability distributions of the zero-energy corner states, and Dirac mass profiles in a circular topological insulator by varying the long-range hopping index $n\!=\!1, 2, 3, 4$ in the $x$ direction, while keeping the hopping index fixed at $m\!=\!1$ in the $y$ direction, as shown in Fig.~\ref{fig:E_OBCxy_n_m1_circle_together}.

Figure~\ref{fig:E_OBCxy_n_m1_circle_together} shows that four zero-energy corner modes (highlighted in red in both the energy levels and probability distributions) persist across all values of $n$. Notably, the spatial positions of these corner states can be tuned by changing the long-range hopping in the $x$ direction. For $n\!=\!m\!=\!1$, which corresponds to the nearest-neighbor hopping case, the four corner states are located at normal angles $\alpha\!=\!\pi/4,3\pi/4,5\pi/4,7\pi/4$, as shown in Fig.~\ref{fig:E_OBCxy_n_m1_circle_together}(a2). Here $\alpha$ denotes the angle between the edge\rq{}s normal direction and the $+x$ axis and is used to define the orientation of the corner states. As the long-range hopping index $n$ increases with $m\!=\!1$ fixed, the locations of the corner modes shift and gradually approach, but do not reach, $\alpha\!=\!\pi/2$ and $3\pi/2$ from opposite directions, as shown in Figs.~\ref{fig:E_OBCxy_n_m1_circle_together}(b2)-\ref{fig:E_OBCxy_n_m1_circle_together}(d2).

\subsection{Case of $m\!=\!1,2,3,4$ with $n\!=\!1$}\label{3_2}

\begin{figure*}[htpb]
\centering
\includegraphics[width=0.65\textwidth]{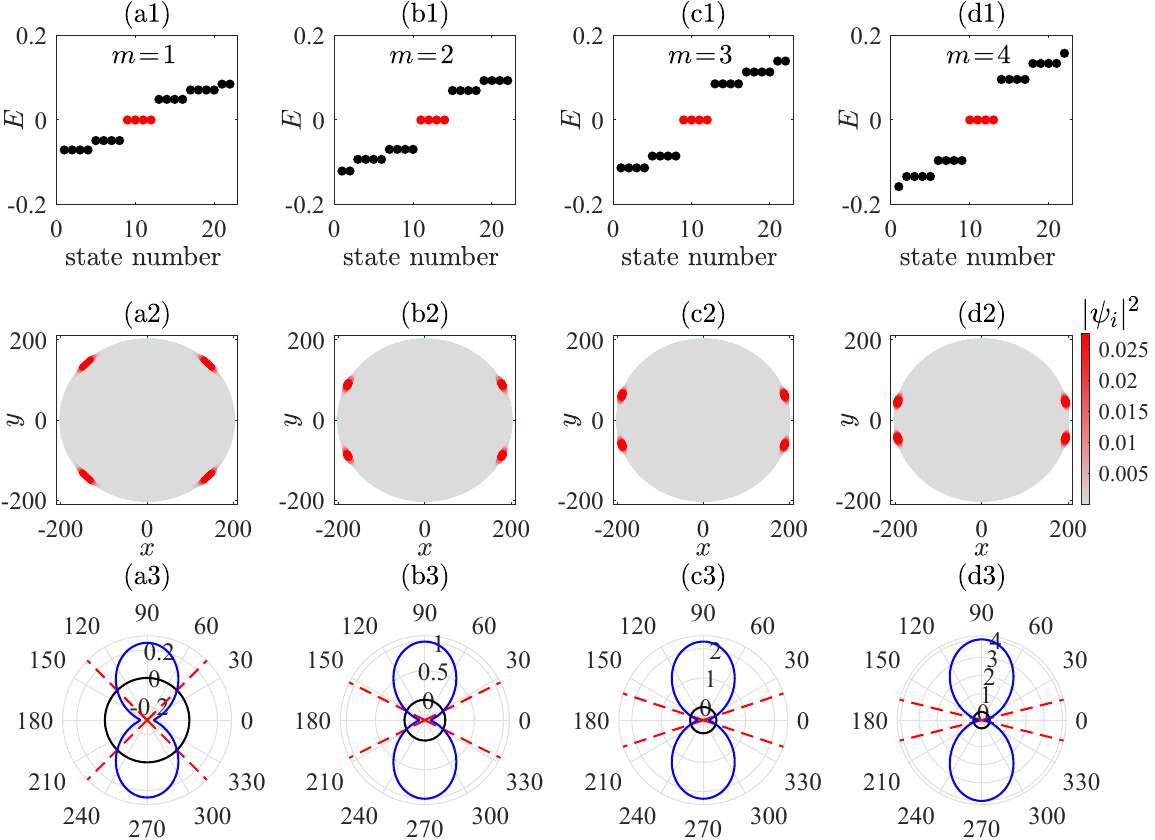}
\caption{Energy levels, corner states, and Dirac mass in a circular topological insulator with varying long-range hoppings in the $y$ direction, i.e., by changing $m\!=\!1,2,3,4$ while keeping $n\!=\!1$ fixed. (a1-d1) Energy levels. Here the red dots correspond to zero-energy modes. (a2-d2) Probability distribution of the corner states, highlighted in red. (a3-d3) Dirac mass (blue curve) [Eq.~\eqref{eq:Dirac_mass}] as a function of the normal angle $\alpha$ in the polar coordinate chart. Here the black circle describes the zeros of Dirac mass along different polar angles, while the two red dashed lines indicate the normal directions corresponding to the corner modes. We set $n\!=\!1$, while all other parameters are kept identical to those in Fig.~\ref{fig:E_OBCxy_n_m1_circle_together}.}
\label{fig:E_OBCxy_n1_m_circle_together}
\end{figure*}

We analyze the energy levels, probability distributions of corner states, and Dirac mass profiles in circular topological insulators with varying long-range hoppings in the $y$ direction by changing $m \!=\!1, 2, 3, 4$, while keeping the hopping index fixed at $n\!=\!1$ in the $x$ direction, as shown in Fig.~\ref{fig:E_OBCxy_n1_m_circle_together}.

As in Sec.~\ref{3_1}, four zero-energy corner modes consistently appear across all values of $m$, and their spatial positions can be tuned by modifying the long-range hopping in the $y$ direction. However, in contrast to the $x$-direction case, the normal angles $\alpha$ corresponding to the corner states now shift toward, but remain slightly away from, $\alpha\!=\!0$ and $\pi$ as $m$ increases. This distinction arises from the directionality of the modified hopping terms. Consequently, tuning long-range hoppings in different directions ($x$ vs $y$) leads to distinct regulatory effects on the spatial localization of the corner modes.

\subsection{Explanation}\label{3_3}

We now explain why the positions of the corner states, characterized by the normal angle $\alpha$ that defines their direction, shift with varying long-range hoppings. These positions correspond to the points where the Dirac mass reverses or vanishes, that is, where the black and blue curves intersect in Figs.~\ref{fig:E_OBCxy_n_m1_circle_together}(a3)-\ref{fig:E_OBCxy_n_m1_circle_together}(d3) and Figs.~\ref{fig:E_OBCxy_n1_m_circle_together}(a3)-\ref{fig:E_OBCxy_n1_m_circle_together}(d3). Furthermore, Sec.~\ref{4} presents the analytical expression for the Dirac mass and discusses its vanishing in relation to the emergence of zero-energy modes.

\section{Edge theory and Dirac mass}\label{4}

To explore the relationship between the vanishing Dirac mass and the emergence of zero-energy modes, we first map the tight-binding model Hamiltonian~\eqref{eq:Hk_TB} to a continuous model, and then derive the effective edge Hamiltonian to analytically obtain the Dirac mass associated with the zero-energy modes within the edge theory.

\subsection{Continuous model}\label{4_1}

In order to map the tight-binding model Hamiltonian~\eqref{eq:Hk_TB} to a continuous model, we utilize the replacements~\cite{qin2022phase,qin2022light,qin2023light,shen2017topological}
\begin{eqnarray}
&&\sin(k_{i}a_{i})\rightarrow k_{i}a_{i},\label{eq:sin}\\
&&\cos(k_{i}a_{i})\rightarrow 1 \!-\! \frac{1}{2}k_{i}^{2}a_{i}^2,\label{eq:cos}
\end{eqnarray} where $i\!=\!x,y,z$ and $a_{i}$ is the lattice constant in the $i$ direction.
By substituting Eqs.~\eqref{eq:sin} and \eqref{eq:cos} into Eq.~\eqref{eq:Hk_TB}, we get a higher-order topological insulator described by the BHZ model with an additional momentum-dependent spin-orbital potential as
\begin{equation}
\hat{\cal H}^{}({\bf k})\!=\!\hat{\cal H}_{\rm BHZ}({\bf k}) \!+\! \hat{\cal H}\rq{}({\bf k}),\label{eq:Hk_0}
\end{equation}
where $\hat{\cal H}_{\rm BHZ}^{}({\bf k})\!=\!(M\!-\!Bk^{2})\sigma_{0}\otimes\tau_{z} \!+\! Ak_{x}\sigma_{z}\otimes\tau_{x} \!+\! Ak_{y}\sigma_{0}\otimes\tau_{y}$ is the BHZ model~\cite{bernevig2006quantum} and $\hat{\cal H}\rq{}({\bf k})\!=\!V({\bf k})\sigma_{x}\otimes\tau_{x}$ represents a momentum-space perturbation spin-orbit-coupling potential term with its momentum distribution determined by $V({\bf k})\!=\!g[(nk_{x})^{2}\!-\!(mk_{y})^{2}]$, with $n,m\in\mathbb{Z}$.
Physically, $\hat{\cal H}_{\rm BHZ}({\bf k})$ supports a pair of gapless helical edge states within the bulk gap~\cite{bernevig2006quantum}.

\subsection{Dirac mass and zero-energy modes}\label{4_2}

We analyze the system described by Eq.~\eqref{eq:Hk_0} in momentum space using edge theory~\cite{chen2024tunable,li2024creation,zhou2008finite,shan2010effective,lu2010massive,qin2022phase,slager2015impurity} to derive the Dirac mass and establish its connection to the zero-energy modes. We focus on states localized along an edge defined by the polar angle $\theta$. To describe such edge states, we perform a coordinate rotation of the momentum space $(k_x, k_y)$ counterclockwise about the origin $(0,0)$ by an angle $\theta$, resulting in new coordinates $(k_{x}^{\prime}, k_{y}^{\prime})$. The transformation between the original and rotated coordinates is given by
\begin{eqnarray}\label{eq:kxky}
\!\left\{ \begin{array}{l}
k_{x}\!=\!k_{x}^{\prime}\cos\theta \!+\! k_{y}^{\prime}\sin\theta,\\
k_{y}\!=\!k_{y}^{\prime}\cos\theta \!-\! k_{x}^{\prime}\sin\theta.
\end{array}\right.
\end{eqnarray}
By substituting Eq.~\eqref{eq:kxky} into the Hamiltonian in Eq.~\eqref{eq:Hk_0}, we obtain the rotated model Hamiltonian in the $(k_{x}^{\prime}, k_{y}^{\prime})$ coordinate system
\begin{equation}
\hat{\cal H}^{}({\bf k}^{\prime},\theta)\!=\!\hat{\cal H}_{\rm BHZ}({\bf k}^{\prime},\theta) \!+\! \hat{\cal H}\rq{}({\bf k}^{\prime},\theta),\label{eq:Hk_1}
\end{equation} where ${\bf k}^{\prime}\!=\!(k_{x}^{\prime},k_{y}^{\prime})$ and
\begin{eqnarray}
\hat{\cal H}_{\rm BHZ}({\bf k}^{\prime},\theta)\!=\!
\begin{pmatrix}
\hat{\cal H}_{\rm BHZ}^{\uparrow}({\bf k}^{\prime},\theta) & 0 \\
0 & \hat{\cal H}_{\rm BHZ}^{\downarrow}({\bf k}^{\prime},\theta)
\end{pmatrix},\label{eq:Hk_BHZ_new}
\end{eqnarray}
\begin{eqnarray}
\hat{\cal H}^{\prime}({\bf k}^{\prime},\theta)
\!=\!\!G({\bf k}^{\prime},\theta)\sigma_{x}\otimes\tau_{x},\label{eq:H_perturbation_kxky_rq}
\end{eqnarray}
with
\begin{eqnarray}
G({\bf k}^{\prime},\theta)&\!\!=\!\!&g\!\left[\frac{1}{2}(n^{2}\!+\!m^{2})(k_{x}^{\prime2}\!-\!k_{y}^{\prime2})\cos(2\theta) \right.\nonumber\\
&&\left.\!+\frac{1}{2}(n^{2}\!-\!m^{2})(k_{x}^{\prime2}\!+\!k_{y}^{\prime2})\!+\!(n\!+\!m)k_{x}^{\prime}k_{y}^{\prime}\sin(2\theta)\!\right]\!,\nonumber\\
\end{eqnarray}
\begin{eqnarray}
\hat{\cal H}_{\rm BHZ}^{\uparrow}({\bf k}^{\prime},\theta)\!=\!
\begin{pmatrix}
M\!-\!Bk^{\prime2} & Ae^{i\theta}(k_{x}^{\prime}\!-\!ik_{y}^{\prime}) \\
Ae^{-i\theta}(k_{x}^{\prime}\!+\!ik_{y}^{\prime}) & -(M\!-\!Bk^{\prime2})
\end{pmatrix},\label{eq:Hk_BHZ_up}\nonumber\\
\end{eqnarray}
\begin{eqnarray}
\hat{\cal H}_{\rm BHZ}^{\downarrow}({\bf k}^{\prime},\theta)\!=\!
\begin{pmatrix}
M\!-\!Bk^{\prime2} & -Ae^{-i\theta}(k_{x}^{\prime}\!+\!ik_{y}^{\prime}) \\
-Ae^{i\theta}(k_{x}^{\prime}\!-\!ik_{y}^{\prime}) & -(M\!-\!Bk^{\prime2})
\end{pmatrix}.\label{eq:Hk_BHZ_down}\nonumber\\
\end{eqnarray}

By replacing $k_{y}^{\prime }\!\rightarrow\! -i\partial_{y}^{\prime }$, we obtain the wave function for edge states on the $y^\prime\!=\!0$ boundary from Eq.~\eqref{eq:Hk_BHZ_new}. Projecting the total Hamiltonian~\eqref{eq:Hk_1} onto this edge-state wave function at $k_{x}^{\prime}\!=\!0$, we derive an effective edge Hamiltonian
\begin{eqnarray}
\hat{\cal H}_{\text{edge}}(k_{x}^{\prime},\theta)\!=\!\begin{pmatrix}
Ak_{x}^{\prime} & e^{i\theta}\Delta(k_{x}^{\prime},\theta) \\
e^{-i\theta}\Delta(k_{x}^{\prime},\theta) & -Ak_{x}^{\prime}
\end{pmatrix}. \label{eq:Hk_edge_kx}
\end{eqnarray}
A detailed derivation of the effective edge Hamiltonian~\eqref{eq:Hk_edge_kx} is provided in Sec. SII of the Supplemental Material~\cite{supp}.
Here the Dirac mass term is
\begin{eqnarray}
\Delta(k_{x}^{\prime},\theta)&\!=\!&g\left[\frac{1}{2}(n^{2}\!-\!m^{2})\left(k_{x}^{\prime2}\!+\!\frac{M}{B}\right) \right.\nonumber\\
&&\left.\!+\frac{1}{2}(n^{2}\!+\!m^{2})\left(k_{x}^{\prime2}\!-\!\frac{M}{B}\right)\cos(2\theta)\right].
\end{eqnarray}
At $k_{x}^{\prime}\!=\!0$, this simplifies to~\cite{li2024creation}
\begin{eqnarray}
\Delta(\theta)\!=\!\frac{gM}{2B}\left[ (n^{2}\!-\!m^{2})\!-\!(n^{2}\!+\!m^{2})\cos(2\theta) \right].\label{eq:Dirac_mass}
\end{eqnarray}
The corresponding eigenenergies of the edge Hamiltonian~\eqref{eq:Hk_edge_kx} are
\begin{eqnarray}
E_{\text{edge}}^{(\pm)}(k_{x}^{\prime},\theta)\!=\!\pm\sqrt{A^{2}k_{x}^{\prime2}\!+\!\left[\Delta(k_{x}^{\prime},\theta)\right]^{2}}.
\end{eqnarray}
In particular, at $k_{x}^{\prime}\!=\!0$, the spectrum exhibits an energy gap of $\Delta_{g}\!=\!|2\Delta_{0}(\theta)|$. This gap closes and reopens at the critical angle satisfying $\cos(2\theta)\!=\!(n^{2}\!-\!m^{2})/(n^{2}\!+\!m^{2})$, indicating that the Dirac mass, and thus the edge gap, depends on the edge orientation.
For the special case $n\!=\!m$, the condition simplifies to $\cos(2\theta)\!=\!0$, so the energy gap closes and reopens at $\theta\!=\!n^{\prime}\pi/4$ with $n^{\prime}\!=\!1,3,5,7$, given that $\theta\in[0,2\pi)$.

\begin{figure}[htpb]
\centering
\includegraphics[width=0.85\columnwidth]{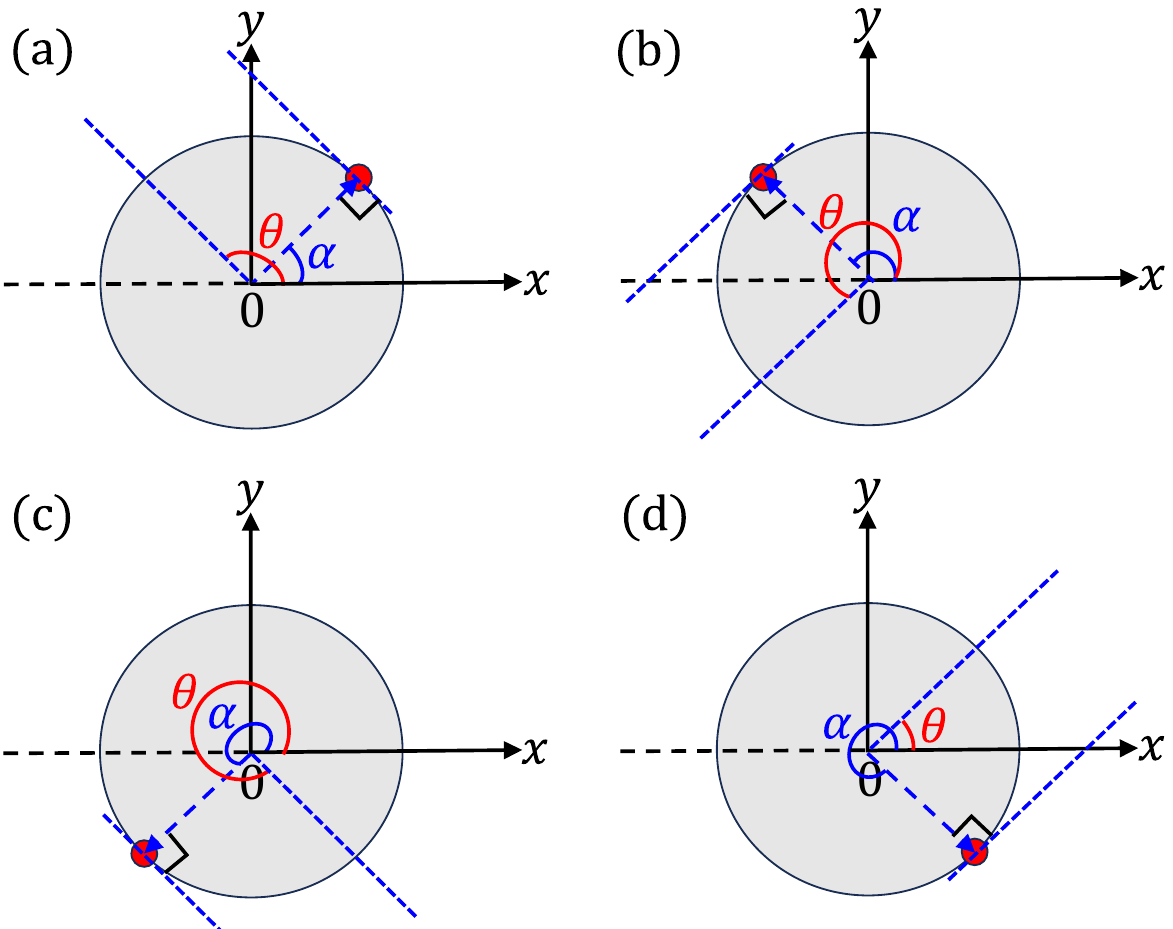}
\caption{Schematic diagram of the normal and tangential angles of the corner states in circular topological insulators. The angle $\alpha$ is the normal angle of the corner state and the angle $\theta$ is the tangential angle of the corner state. The red dot at the tangential edge denotes the position of the corner state so that the normal angle $\alpha$ defines the direction of the corner state. (a) Corner state located in the first quadrant with $\alpha\!=\!\theta\!-\!\frac{\pi}{2}$. (b) Corner state located in the second quadrant with $\alpha\!=\!\theta\!-\!\frac{\pi}{2}$. (c) Corner state located in the third quadrant with $\alpha\!=\!\theta\!-\!\frac{\pi}{2}$. (d) Corner state located in the fourth quadrant with $\alpha\!=\!\theta\!+\!\frac{3\pi}{2}$.}
\label{fig:Schematic_Diagram}
\end{figure}

To verify that only the tangential edges, where the corner states reside, exhibit gapless energy bands or zero Dirac mass, we analyze a circular topological insulator with edge orientations characterized by different normal angles $\alpha$ and tangential angles $\theta$, as illustrated in Fig.~\ref{fig:Schematic_Diagram}. Here, $\alpha$ denotes the angle between the edge's normal direction and the $+x$ axis, while $\theta$ represents the angle between the edge's tangential direction and the $+x$ axis. These angles are related by $\alpha\!=\!\theta\!-\!\frac{\pi}{2}\!+\!2N\pi$ with $N\in\mathbb{N}\!=\!\{0,1,2,3,\ldots\}$.

\section{Geometry-dependent tunable corner states}\label{5}

\begin{figure*}[htpb]
\centering
\includegraphics[width=0.9\textwidth]{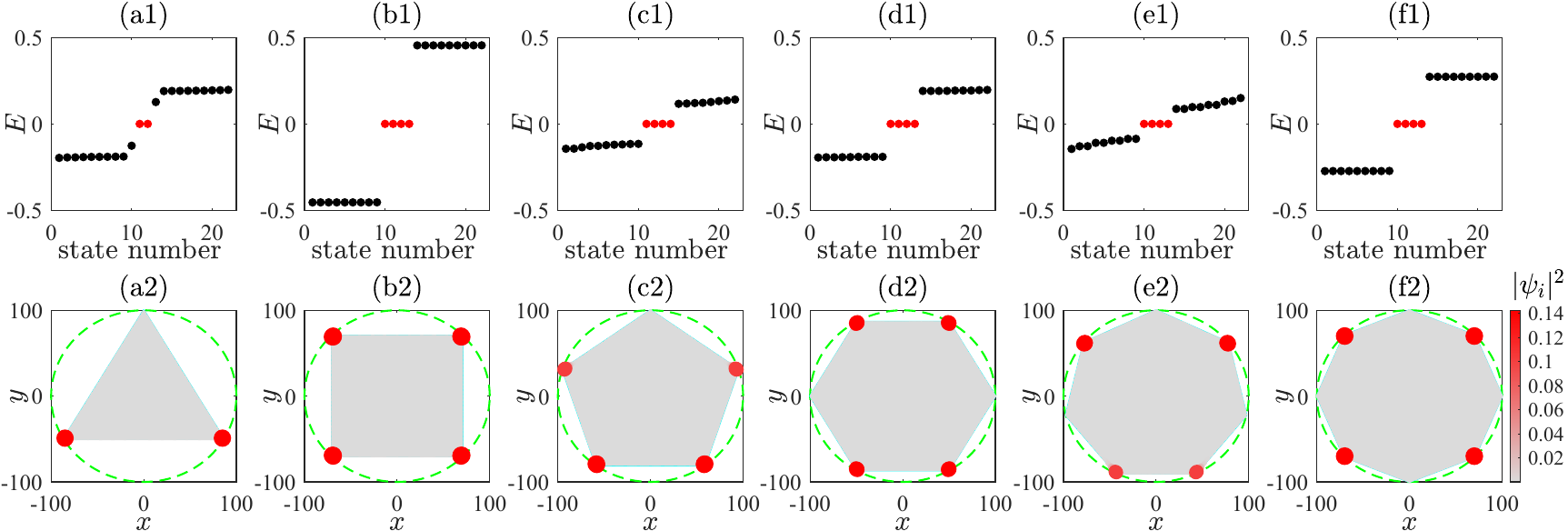}
\caption{Energy levels and corner states in geometrically distinct topological insulators, including triangular, square, pentagonal, hexagonal, heptagonal, and octagonal shapes. (a1-f1) Energy levels. Here the red dots correspond to zero-energy modes. (a2-f2) Probability distribution of the corner states, highlighted in red. We set $n\!=\!m\!=\!2$ and $L_{x}\!=\!L_{y}\!=\!201a$, while keeping all other parameters the same as in Fig.~\ref{fig:E_OBCxy_n_m1_circle_together}.}
\label{fig:E_OBCxy_n2_m2_polygon_together}
\end{figure*}

We present numerical results and detailed discussions about the energy levels and tunable corner states in topological insulators with various geometric shapes, including triangular, square, pentagonal, hexagonal, heptagonal, and octagonal geometries, as illustrated in Fig.~\ref{fig:E_OBCxy_n2_m2_polygon_together}.

Figure~\ref{fig:E_OBCxy_n2_m2_polygon_together} shows that for square and octagonal topological insulators, where the long-range hopping indices are equal in both the $x$ and $y$ directions ($n\!=\!m$), four corner modes emerge. These modes are located at normal angles $\alpha\!=\!\pi/4, 3\pi/4, 5\pi/4, 7\pi/4$, as depicted in Figs.~\ref{fig:E_OBCxy_n2_m2_polygon_together}(b2) and \ref{fig:E_OBCxy_n2_m2_polygon_together}(f2). This behavior is explained by the vanishing of the Dirac mass, i.e., solutions of $\Delta(\theta)\!=\!0$ in Eq.~\eqref{eq:Dirac_mass}, with $\theta\!=\!\alpha\!+\!\frac{\pi}{2}\!-\!2N\pi$ under the conditions $\theta,\alpha\in[0,2\pi)$ and $N\in\mathbb{N}$.

In contrast, the corner modes in other geometries do not appear at these angles. For instance:
\begin{itemize}
\item In the triangular case, the corner states appear at $\alpha\!=\!7\pi/6$ and $11\pi/6$, as shown in Fig.~\ref{fig:E_OBCxy_n2_m2_polygon_together}(a2).

\item In the pentagonal geometry, the corner states are found at $\alpha\!=\!\pi/10,9\pi/10,13\pi/10,17\pi/10$ [Fig.~\ref{fig:E_OBCxy_n2_m2_polygon_together}(c2)].

\item In the hexagonal case, the corner states are located at $\alpha\!=\!\pi/3,2\pi/3,4\pi/3,5\pi/3$ [Fig.~\ref{fig:E_OBCxy_n2_m2_polygon_together}(d2)].

\item In the heptagonal geometry, the corner states occur at $\alpha\!=\!3\pi/14,11\pi/14,19\pi/14,23\pi/14$ [Fig.~\ref{fig:E_OBCxy_n2_m2_polygon_together}(e2)].
\end{itemize}
These results clearly indicate that the location of corner states strongly depends on the geometry of the topological insulator.

The origin of these shifts in the corner mode locations can also be understood through the behavior of the Dirac mass in Eq.~\eqref{eq:Dirac_mass}. For example, in the triangular-shaped case, the Dirac mass along the right boundary with angle $\theta\!=\!2\pi/3$ is $\Delta\!=\!-2\cos(4\pi/3)\!=\!1\!>\!0$, while that along the bottom boundary at $\theta\!=\!0$ or $\pi$ is $\Delta\!=\!-2\!<\!0$. The sign change between these two boundaries implies that their intersection, i.e., the corner between them, must undergo a Dirac mass sign reversal, leading to the appearance of a localized corner mode. Similarly, the left corner exhibits a corner mode due to the opposite signs of the Dirac masses along the left [$\theta\!=\!\pi/3$ and $\Delta\!=\!-2\cos(2\pi/3)\!=\!1\!>\!0$] and bottom boundaries. However, the top corner, bounded by the left and right edges, both having positive Dirac mass, does not support a corner mode, since there is no sign change across the adjacent boundaries. Similar reasoning can be applied to the other geometries to explain the observed corner modes.

In summary, we conclude that a corner hosts a localized corner mode when the Dirac masses of its two adjacent edges have opposite signs. Conversely, if the Dirac masses share the same sign, no corner state is present.

\section{Potential experimental realization}\label{6}

We explore potential experimental realizations of our tight-binding lattice model. As noted in the Introduction, a variety of physical platforms have been proposed for implementing corner states in HOTIs. For illustrative purposes, we focus on the realization of our model using electrical circuit systems.

A versatile theoretical framework has been proposed for realizing arbitrary tight-binding lattice models using electrical $LC$ circuits~\cite{dong2021topolectric}. In this scheme, each lattice site is represented by an electrical node, which is connected not only to its nearest neighbors but also to long-range nodes and grounded through appropriately chosen capacitors and inductors. In particular, by extending each node to include $n$ subnodes, each associated with current and voltage phases corresponding to the $n$th roots of unity, it is in principle possible to implement arbitrary hopping amplitudes. These are achieved through carefully designed shift-capacitor couplings between the subnodes.

\section{Conclusion}\label{7}

We have presented a theoretical framework for controlling corner modes in HOTIs with long-range hoppings and diverse geometries. Our study demonstrated that the spatial positions of corner states can be precisely tuned by adjusting the strength and direction of long-range hoppings, as confirmed by edge theory analysis and the condition of vanishing Dirac mass. In circular HOTIs, such tunability allows for smooth manipulation of corner mode locations, while in polygonal geometries, the presence or absence of corner states is determined by the relative signs of Dirac masses on adjacent edges. These results offer a versatile approach for engineering reconfigurable HOTIs and open possibilities for the design of tunable topological materials.

\begin{acknowledgments}
F.Q. acknowledges support from the Jiangsu Specially-Appointed Professor Program in Jiangsu Province and the Doctoral Research Start-Up Fund of Jiangsu University of Science and Technology. R.C. acknowledges support from the National Natural Science Foundation of China (Grant No. 12304195), the Chutian Scholars Program in Hubei Province, the Hubei Provincial Natural Science Foundation (Grant No. 2025AFA081), and the Original Seed Program of Hubei University.
\end{acknowledgments}

\section*{Data Availability}
The data that support the findings of this article are available from the authors upon reasonable request.

%
%
%
\twocolumngrid
\bibliography{references_Higher_2D}

\end{document}